\documentstyle[aps,prl,multicol,fancyheadings,epsfig,amsbsy,amssymb,amstex]{revtex}

\def\bbbc{{\mathchoice {\setbox0=\hbox{$\displaystyle\rm C$}\hbox{\hbox 
to0pt{\kern0.4\wd0\vrule height0.9\ht0\hss}\box0}} 
{\setbox0=\hbox{$\textstyle\rm C$}\hbox{\hbox 
to0pt{\kern0.4\wd0\vrule height0.9\ht0\hss}\box0}} 
{\setbox0=\hbox{$\scriptstyle\rm C$}\hbox{\hbox 
to0pt{\kern0.4\wd0\vrule height0.9\ht0\hss}\box0}} 
{\setbox0=\hbox{$\scriptscriptstyle\rm C$}\hbox{\hbox 
to0pt{\kern0.4\wd0\vrule height0.9\ht0\hss}\box0}}}} 
\begin{document} 
\draft 
%inerst title 
\title{Itinerant Ferromagnetism in the Periodic Anderson Model} 
\author{C. D. Batista,$^1$ J. Bon\v ca,$^2$ and J. E. Gubernatis$^1$} 
\address{$^1$Center for Nonlinear Studies and Theoretical Division\\ 
Los Alamos National Laboratory, Los Alamos, NM 87545\\ 
$^2$ Department of Physics, FMF\\ University of Ljubljana and J. Stefan  
Institute, Ljubljana, Slovenia} 
\date{\today} 
\maketitle 
\begin{abstract} 

We introduce a novel mechanism for itinerant ferromagnetism, 
based on a simple two-band model. The model includes an uncorrelated
and dispersive band hybridized with a second band which is narrow and
correlated.  The simplest Hamiltonian containing these ingredients is
the Periodic Anderson Model (PAM). Using quantum Monte Carlo and
analytical methods, we show that the PAM and an extension of it
contain the new mechanism and exhibit a non-saturated ferromagnetic
ground state in the intermediate valence regime.  We propose that the
mechanism, which does not assume an intra atomic Hund's coupling, is
present in both the iron group and in some $f$ electron compounds like
Ce(Rh$_{1-x}$Ru$_x$)$_3$B$_2$ , La$_x$Ce$_{1-x}$Rh$_3$B$_2$ and the
uranium monochalcogenides US, USe, and UTe.

\end{abstract} 
\pacs{} 
 
%\vspace*{-0.4cm} 
\begin{multicols}{2} 
 
\columnseprule 0pt 
 
\narrowtext 
%\vspace*{-0.5cm} 
% body of paper here 
\section{Introduction}

% Historical prespective

Itinerant ferromagnetism was the first
collective quantum phenomena considered as a manifestation 
of the strong Coulomb interactions which are 
present in an electronic system. However, the origin of this
phenomenon is still an open problem. Here we introduce  a novel mechanism 
for itinerant ferromagnetism, which is based on a simple two-band 
model. This work is an extension of a previous letter \cite{ours1} where we 
have described the basic ideas. In this paper, we show that the mechanism is 
supported by the numerical results obtained from quantum Monte Carlo (QMC) simulations of 
the periodic Anderson model (PAM). We also analyse the experimental consequences for 
some $f$ electron compounds and the Iron group. However, before describing 
the details of our mechanism, it is useful to develop a historical 
perspective for the itinerant ferromagnetism.

Seventy  four years ago, Heisenberg \cite{Heisenberg} formulated 
his spin model to address the problem of ferroamgnetism,
but as Bloch \cite{Bloch} pointed out, a model of localized spins cannot 
explain the metallic ferromagnetism observed in iron, cobalt and nickel.
By introducing the effects of exchange terms into the 
Sommerfeld description of the free-electron gas, Bloch predicted a 
ferromagnetic ground state for a sufficiently low electron density.
In 1934, Wigner \cite{Wigner} added the correlation terms to
the model considered by Bloch. After analyzing the effects of 
the correlations, he concluded
that the conditions for ferromagnetism 
in the electron gas were so stringent as to be never satisfied.  

The first attempt at analyzing a real FM metal, like Ni, was made 
by Slater \cite{Slater}. He concluded that the main contribution to
the exchange energy is provided by intra-atomic interactions.
In the meantime, Stoner \cite{Stoner}
introduced his picture where the metallic ferromagnetism results from holes in the 
$3d$ band interacting via an exchange energy proportional to the relative magnetization and 
obeying Fermi-Dirac statistics.  
However, the model considered first by Stoner \cite{Stoner} 
and later by Wohlfarth \cite{Wohlfarth}, did not take into account the
correlations of the $3d$ electrons, except for the constraints
imposed by the Pauli exclusion principle. In other words, they did not 
consider the fact that the Coulomb repulsion tends to keep the electrons apart.

In 1953, the importance of these correlations was pointed out by van Vleck \cite{Van Vleck}.
He emphasized that the energy required to tear off an electron increases rapidly 
with the degree of ionization. (The energy of two Ni atoms in a $3d^9$ configuration
is appreciably lower than having one atom in the $3d^8$ state and the other one in $3d^{10}$.)
Based on this observation, he proposed an alternative picture (minimum polarity model) where the 
states of higher ionization in Ni are ruled out completely, and the
configuration $3d^{9.4}$ is considered to be 40 percent  $3d^{10}$ and sixty
percent $3d^{9}$. The lattice sites occupied by $3d^{9}$ and $3d^{10}$ configurations
are continuously redistributing in his picture. 
The van Vleck proposal is the precursor of the Hubbard model for infinite $U$.

Following Slater \cite{Slater2}, van Vleck \cite{Van Vleck} speculated that the 
contamination by states of higher polarity, not included in his model, provides 
the exchange interaction (intra-atomic in this case) necessary for ferromagnetism. 
Hence he calculated the effective nearest-neighbor magnetic interaction induced
by second order perturbative fluctuations from the $d^9+d^9$ configuration to the $d^8+d^{10}$.
In this way, van Vleck arrived at a model which describes itinerant and 
correlated (only allowing $d^9$ and $d^{10}$ configurations) holes with 
a nearest-neighbor exchange interaction $\alpha {\bf S_i.S_j}$ (generalized
Heisenberg model). However,
as van Vleck explained at the end of his paper \cite{Van Vleck}, the sign 
and the magnitude of $\alpha$ are very sensitive to the precise values of the 
energies of the different possible intermediate states (singlets or triplets) 
in the $d^8$ configuration.

In 1963, the one-band Hubbard model was proposed independently by Gutzwiller
\cite{Gutzwiller}, Hubbard \cite{Hubbard} and Kanamori \cite{Kanamori} to 
explain the metallic ferromagnetism in the $3d$ transition metals. 
The Hubbard model incorporates the kinetic energy in a
{\it single nondegenerate band} with an intra-atomic Coulomb repulsion
$U$ to describe the electrons in the $s$ band of the transition metals.
In contrast to the previous models,
the Hubbard model does not include any explicit 
exchange interaction which favors a ferromagnetic phase. The 
implicit question raised by this proposal is: Can ferromagnetism
emerge from the interplay between the kinetic energy and 
the Coulomb repulsion, or it is strictly neccesary to include an explicit exchange
interaction provided by the intra-atomic Hund's coupling? 
This simple question becomes even more relevant if we consider  
$f$-electron itinerant ferromagnets, 
like CeRh$_3$B$_2$ \cite{Cornelius}, whose only local magnetic 
coupling is antiferromagnetic.

Unfortunately, 
with the exception of Nagaoka's \cite{Nagaoka} and Lieb's \cite{Lieb,Lieb2} 
theorems, the subsequent theoretical approches were not controlled enough to
determine whether the Hubbard model has a FM phase.
The central issue is the precise evaluation of the
energy for the paramagnetic (PM) phase. Because it does not 
properly incorporate the correlations, mean field theory
overestimates this energy and predicts a large FM region \cite{Vollhardt}. In
contrast, numerical calculations have narrowed the extent of
this phase to a small region around the Nagaoka point \cite{Nagaoka}.

There are several controversies in the history of itinerant ferromagnetism.
Most of them originate in the lack of reliable methodologies to
solve strongly correlated Hamiltonians. In addition, the number of
compounds exhibiting ferromagnetism has increased during the last several
decades. There is a new list of $Ce$ and $U$ based compounds
which are itinerant ferromagnets \cite{santini} with large Curie temperatures
(of the order of 100$^{\circ}$K) and the only explicit magnetic interaction is 
antiferromagnetic (Kondo like). This observation
stimulated us to reconsider the situation of the iron group and ask 
whether the ferromagnetism originates in the intra-atomic 
Hund's interaction or is just a consequence of the strong Coulomb 
repulsion between electrons with a particular band structure. One possible and 
plausible alternative is that these two phenomena are cooperating 
to stabilize a ferromagnetic ground state. In any case, it is important
to elucidate whether a realistic lattice model only containing 
repulsive terms (for instance, the Hubbard model) can sustain 
a ferromagnetic state, or it is necessary to invoke additional terms 
which are explicitly ferromagnetic. The Hubbard Hamiltonian is 
in the first category. However, the most accurate numerical calculations seem 
to indicate that this model does not sustain a ferromagnetic 
ground state. 

%-Present status and challenge

Going beyond the simple one-band Hubbard model
has been advocated, for instance, by Vollhardt {\it et al} \cite{Vollhardt}.
They note that the inclusion of additional Coulomb
density-density interactions, correlated hoppings, and direct exchange
interactions favors FM ordering in the single-band Hubbard model. 
In fact, a very simple analysis
shows that increasing the density of
states ${\cal D}(E)$ below the Fermi energy $E_F$ and placing $E_F$
close to the lower band edge increases the FM tendency. One can
achieve this by including 
a next nearest neighbor hopping $t^\prime$ or by placing the hoppings on
frustrated (non-bipartite) lattices. The effectiveness of 
$t^\prime$ was studied numerically by Hlubina et. al.
\cite{Hlubina} for the 
Hubbard model on a square lattice. They found a FM state when the van Hove
singularity in ${\cal D}(E)$ occurred at $E_F$. However, this
phase was not robust against very small changes in $t^\prime$.

After seven decades of intense effort, the microscopic mechanisms
driving the metallic FM phase are still unknown \cite{Vollhardt,Fazekas}. 
We still do not know what is the minimal lattice model of itinerant 
ferromagnetism and, more importantly,
the basic mechanism of ordering.
%- Why the Periodic Anderson Model 

While the Hubbard model is so reluctant to have a FM state, there is
an increasing amount of evidence indicating that the Periodic Anderson 
Model (PAM) has a FM phase in a large region of its quantum phase diagram 
\cite{ours1,Bonca,ours2,Guerrero,Guerrero2,Moller,sbmft2,sbmft3,Tahvildar-Zadeh,meyer1,meyer2,meyer3}.
Since the $d$ orbitals of the transition metals are hybridized with
the $s-p$ bands, we can consider the inclusion of a second band
as the next step in the search of itinerant ferromagnetism
from pure Coulomb repulsions. Indeed a very simple extension of the PAM
can be used to describe the physics of the iron group \cite{Nolting,ours2}.
In addition, there is large number of cerium and uranium compounds, like 
Ce(Rh$_{1-x}$Ru$_x$)$_3$B$_2$ ($T_c=115K$) \cite{malik,Berger}, CeRh$_3$B$_2$ \cite{Cornelius}, 
and US, USe, and UTe \cite{santini} which are metallic ferromagnets and can be 
describred with the PAM.

%- DMFT

Ferromagnetism is readily found in the PAM by mean-field approximations 
in any  dimensions \cite{Moller,sbmft2,sbmft3,Tahvildar-Zadeh,meyer1,meyer2,meyer3}. 
Using a slave-boson mean-field theory (SBMFT) for the symmetric PAM, 
M\"oller and W\"olfe \cite{Moller} found  a PM or antiferromagnetic (AF) 
phase at half filling depending on the value of the Coulomb repulsion $U$. By lowering 
the density of electrons from 1/2 filling, they also found a smooth 
crossover from AF to FM order via a spiral phase.  Just before 1/4 
filling, they got a first-order transition from FM to AF order. More 
recently, the SBMFT calculations of Doradzi\' nski and Spalek 
\cite{sbmft2,sbmft3} found wide regions of ferromagnetism in the intermediate 
valence regime that surprisingly extended well below 1/4 filling. 
 
A ferromagnetic phase is also obtained when the 
dynamical mean-field theory (DMFT) is applied to the PAM
\cite{Tahvildar-Zadeh,meyer1,meyer2,meyer3}.
Tahvildar-Zadeh {\it et al.\/} found a region of ferromagnetism and studied its temperature 
dependence.  At very low temperatures, their ferromagnetic region 
extended over a wide range of electron fillings and in many cases 
embraced the electron filling of 3/8. They proposed a specific 
Kondo-induced mechanism for ferromagnetism at 3/8 filling that has the 
conduction electrons in a spin-polarized charge density-wave 
anti-aligned with the ferromagnetically aligned local moments on the 
valence orbitals. More recently Meyer and 
Nolting\cite{meyer1,meyer2,meyer3} appended perturbation theory to DMFT 
and also predicted ferromagnetism over a broad range of electron 
filling extending below 1/4 filling. In addition Schwieger and Nolting \cite{Schwieger} also
considered an extension of the PAM, similar to the one considerd here, 
to estimate the importance of $s-d$ hybridization for the magnetic properties of 
transition metals. 

%- DMRG  

There is also a considerable amount of numerical evidence
showing ferromagnetic solutions for the ground state of the PAM. 
Noack and Guerrero \cite{Guerrero}, for example, found 
partially and completely saturated ferromagnetism
using the density matrix renormalization group (DMRG) method in 
one dimension.  They 
considered a parameter regime where there is one electron in each $f$ orbital. 
For a sufficiently large value of $U$, the 
model exhibited a ferromagnetic ground state. Beyond an 
interaction-dependent value of the doping and a doping-dependent value 
of $U$, this state disappeared. The ferromagnetic 
phase was a peninsula in a phase diagram that was otherwise a sea of 
paramagnetism except at 1/4 and 1/2 filling where the ground state of 
the PAM was antiferromagnetic.

Our previous \cite{Bonca} and new QMC results qualitatively agree with the DMRG 
work; however, the phases we find quantitatively and qualitatively 
disagree with those derived from the mean-field approximations. 
Quantitatively, we find ferromagnetism in a narrower doping range than 
the one predicted by the DMFT and SBMFT calculations.  For fillings 
between 3/8 and 1/2, QMC predicts a PM region, whereas 
mean-field theory predicts ferromagnetic states in part of that 
region. In fact, at a filling of 3/8 where DMFT calculations predict 
ferromagnetism, we find a novel ground state of an entirely different 
symmetry. Instead of ferromagnetism, QMC finds a resonating spin 
density-wave (RSDW) state; that is, the ground state was a linear 
combination of two degenerate spin-density waves characterized by the 
$(\pi,0)$ and $(0,\pi)$ wave vectors.

The novel mechanism we introduce in the present paper operates 
when the system is in a mixed valence regime. This regime has 
been studied numerically only in the context of DMFT \cite{meyer1}. 
We will show however that the ferromagnetic solution obtained 
with DMFT in the mixed valence regime has a different origin
and therefore is not representative of our new mechanism.
The main ingredients for our mechanism are an uncorrelated dispersive
band which is hybridized with a correlated and narrow band.  
We show the PAM supports our mechanism by doing 
quantum Monte Carlo (QMC) simulations on one and two 
dimensional lattices. The results of these simulations are interpreted 
with an effective Hamiltonian derived from the PAM. In this way, we
establish that the new mechanism can be interpreted as a
generalization to the lattice of the first Hund's rule for the atom. 
The two level 
band structure generated by the gap recreates for the lattice, the
shell-like level strucure of the hydrogenic atom.
When the lower shell is incomplete, the local part of 
the Coulomb interaction is minimized by polarizing the electrons
which are occupying the incomplete shell.

\section{Model}
 
%- Periodic Anderson Model

%- Extension of PAM to describe transition metals. 

The PAM was originally introduced to explain the properties of the 
the rare-earth and actinide metallic compounds including the so called 
heavy fermion compounds. A very simple extension of this model can also 
be applied to the description of many transition metals \cite{Fulde,Nolting}.
The basic ingredients of this model are a narrow and 
correlated $f$ band hybridized with a despersive and uncorrelated $d$ band.
The Hamiltonian associated with this model is:

\begin{eqnarray}
H=H_0+H_U 
\nonumber
\end{eqnarray}

\begin{eqnarray}
H_0 =- &t_{a}& \sum_{\langle {\bf r,r'} \rangle,\sigma} (a_{{\bf r}\sigma}^\dagger
  a^{}_{{\bf r'}\sigma}+ a_{{\bf r'}\sigma}^\dagger
  a^{}_{{\bf r}\sigma})  + \epsilon_a\sum_{{\bf r},\sigma}
   n_{{\bf r}\sigma}^a 
\nonumber \\
 -&t_b&\sum_{\langle {\bf r,r'} \rangle,\sigma} (b_{{\bf r}\sigma}^\dagger
  b^{}_{{\bf r'}\sigma}+ b_{{\bf r'}\sigma}^\dagger b^{}_{{\bf r}\sigma})
\nonumber \\
  + &V& \sum_{{\bf r},\sigma} (b_{{\bf r}\sigma}^\dagger
  a^{}_{{\bf r}\sigma}+a_{{\bf r}\sigma}^\dagger
  b^{}_{{\bf r}\sigma}), 
\nonumber
\end{eqnarray}

\begin{eqnarray}
H_U= \frac{U}{2}
  \sum_{{\bf r},\sigma}n_{{\bf r}\sigma}^an_{{\bf r}\bar {\sigma}}^a\ ,
\end{eqnarray}
where $b_{{\bf r}\sigma}^\dagger$ and $a_{{\bf r}\sigma}^\dagger$ create an
electron with spin $\sigma$ in $b$ and $a$ orbitals at lattice
site ${\bf r}$ and $n^a_{{\bf r}\sigma}=a^{\dagger}_{{\bf r}\sigma}a^{}_{{\bf r}\sigma}$.
The $t_b$ and $t_a$ hoppings are only to
nearest-neighbor sites. When $t_a=0$, the Hamiltonian is the
standard PAM. For the $f$ electron compounds, the $a$ and $b$ orbitals  
play the role of the $f$ and $d$ orbitals, and $t_a\approx 0$. For transition metals,
they correspond to the $3d$ and $4s$ orbitals. Unless otherwise specified, we will set
$t_b=1$.

For $U=0$, the resulting 
Hamiltonian $H_0$ is easily diagonalized:
\begin{eqnarray}
H_{0} = \sum_{{\bf k},\sigma}
\left ( E_{{\bf k}}^{+} \beta^{\dagger}_{{\bf k}\sigma}
\beta^{\;}_{{\bf k}\sigma} + E_{{\bf k}}^{-} \alpha^{\dagger}_{{\bf k}\sigma}
\alpha^{\;}_{{\bf k}\sigma} \right )
\end{eqnarray}
where the dispersion relations for the upper and the lower bands are:
\begin{eqnarray}
E_{{\bf k}}^{\pm}=\frac{1}{2} \Biggl[
e_{\bf k}^b+e_{\bf k}^a \pm \sqrt{(e_{\bf k}^b -e_{\bf k}^a)^2+4V^2} \Biggr],
\end{eqnarray}
with 
\begin{eqnarray}
e_{\bf k}^b &=& -2t_b \sum^D_{i=1} \cos k_{x_i},
\nonumber \\
e_{\bf k}^a &=& \epsilon_a-2t_a \sum^D_{i=1} \cos k_{x_i},
\end{eqnarray}
for a hypercubic 
lattice in dimension $D$. The operators
which create quasi-particles in the lower and upper bands are:

\begin{eqnarray}
 \alpha_{{\bf k}\sigma}^\dagger
    &=&  u^{}_{{\bf k}} a_{{\bf k}\sigma}^\dagger
                     +v^{}_{{\bf k}} b_{{\bf k}\sigma}^\dagger, 
\nonumber \\
 \beta_{{\bf k}\sigma}^\dagger
    &=&  -v^{}_{{\bf k}} a_{{\bf k}\sigma}^\dagger
                     +u^{}_{{\bf k}}  b_{{\bf k}\sigma}^\dagger,
\label{quasi}
\end{eqnarray}
with
\begin{eqnarray}
u_{\bf k}&=& \frac{E^+_{\bf k}-e_{\bf k}^a}{\sqrt{(E^+_{\bf k}-e_{\bf k}^a)^2+V^2}}, 
\nonumber \\
v_{\bf k}&=& \frac{-V} {\sqrt{(E^+_{\bf k}-e_{\bf k}^a)^2+V^2}}.
\label{alphak}
\end{eqnarray}
The noninteracting bands $E_{{\bf k}}^{\pm}$ are plotted in Fig.~1 for a one 
dimensional system. If $|V| \ll |t_b|$, we can identify regions with well defined 
$f$ or $d$ character in the lower and the upper bands. In particular, the case
illustrated in Fig.~1 corresponds to a situation where the $a$ and the 
$b$ bands were crossing before being hybridized ($\epsilon_a > -2t_b$). We 
see a small region in the center of the lower band which is dispersive 
and large regions on both sides which are nearly flat. The upper band
exhibits the opposite behavior. The nearly flat regions in each band correspond 
to states with a predominant $f$ character, while the dispersive regions are
associated to the states with $d$ character. 

\begin{figure}[tbp]
\begin{center}
\vspace{-1.0cm} \epsfig{file=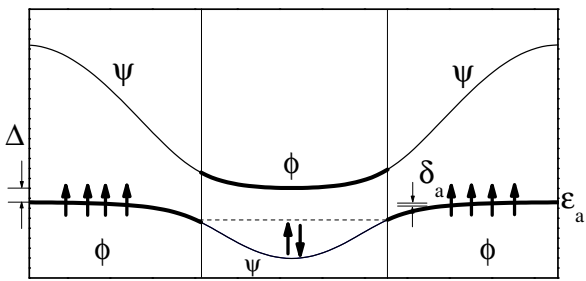,height=60mm,angle=-90}
\end{center}
\vspace{-1.5cm} \caption{Illustration of the effective model and
the FM mechanism. $\Delta$ is the hibridization gap and $\delta_a$
is the interval of energy where the electrons are polarized.} 
\label{fig1}
\end{figure}

\section{New Mechanism for Ferromagnetism} 

%-Introduction to the known regimes

%- This mechanism is not double-exchange

%- Definition of the intermediate Valence Regime

%- Why a non-perturbative approach is required?

%- Derivation of the effective model.

%- Description of the mechanism.

%- Comparison with Hund's rule in the atomic case.
 
The PAM has different regimes depending on the 
values of its parameters and the particle concentration $n=N_e/4N$
($N_e$ is the total number of particles and $N$ is the number of unit cells). 
If $V \ll |\epsilon_a|$ and $V \ll |U+ \epsilon_a|$, there is one particle (magnetic moment) localized
in each $a$ orbital, and the fluctuations to the conduction band can 
be considered in a perturbative way. By this procedure, the PAM can be
reduced to the Kondo Lattice Model (KLM) \cite{Schrieffer} which contains only
one parameter $J_K/t$ (with $J_K=$ and $t=t_b$). The KLM 
has been extensively studied \cite{Doniach0,Jullien,Solyom,Muller}, 
and the evolution of its phase diagram is described for instance in 
a review article by Tsunetsugu {\it et al} \cite{Tsunetsugu}.
One of the earliest approaches to the KLM is the mean field treatment of
Doniach \cite{Doniach0} for the related one dimensional Kondo necklace. For half filling,
this approximation leads to a transition from a N\'eel ordered 
state in the weak coupling regime ($J_K \ll |t|$) to a nonmagnetic `Kondo singlet' state 
above the critical value $J_K^c = t$. 

Lacroix and Cyrot \cite{Lacroix} did a more extensive 
mean field treatment for three dimensional KLM. They also found a magnetically ordered
state for weak coupling. For low density of conduction electrons. In their phase diagram,  
the ordered state is ferromagnetic  for low and intermediate densities of conduction electrons,
and antiferromagnetic in the vicinity of half filling. The `Kondo singlet' phase appears above 
some critical value of $J_K^c(n)$ in the whole range of concentrations. 

Using another mean field treatment for the one dimensional KLM, Fazekas 
and M\"uller-Hartmann \cite{Muller} obtained a phase diagram containing 
only magnetically ordered phases: spiral below some critical value of $J_K/t$
which depends on the particle density and ferromagnetic above this value. 
To get this result, they fixed the orientation of the localized spins in 
a spiral ordering and minimized the total energy with respect to the 
wave vector of the spiral. Even though this treatment of the spin polarized 
state is valid for classical spins, it neglects completely the Kondo singlet
formation which occurs in the strong coupling limit for the considered case ($S=1/2$).

Sigrist {\it et al} \cite{Sigrist2} gave an exact treatment of one 
dimensional KLM for the strong coupling regime $J_K \gg t$ finding a ferromagnetic
phase for any particle density. However, it is important to remark that 
the mechanism driving the ferromagnetism in the later case is not the same 
as the double-exchange mechanism associated with the mean field solution of Fazekas 
and M\"uller-Hartmann \cite{Muller}. To understand this difference, we just need to
notice that for $J_K/t=\infty$ mean field \cite{Muller} predicts a 
ferromagnetic solution 
while the exact solution has a complete spin degeneracy. Therefore 
double-exchange is not the mechanism driving the ferromagnetic phase 
of the KLM (at least when the localized spins are $S=1/2$). 

The real mechanism has
been unveiled by Sigrist {\it et al} \cite{Sigrist2} who used degenerate perturbation
theory to determine the lifting of this degeneracy when the ratio $J_K/t$ becomes finite. 
The new ground state is an itinerant ferromagnet for any concentration of conduction 
electrons. In this state,  the spins which are not participating
in  Kondo singlet are fully polarized. We can see from their solution that the
motion of the Kondo singlets stabilizes the FM state in a way similar to 
Nagaoka's solution \cite{Nagaoka}. The second order 
effective Hamiltonian obtained after the perturbative calculation includes 
nearest-neighbor hopping $t/2$, plus a next-nearest-neighbor 
correlated hopping $t'$ which is order $t^2/J_K$. Then there are two different ways
to move a Kondo singlet from one site to its next-nearest-neighbor: by
two applications of $t/2$ or by one application of $t'$. Only when the background is
ferromagnetic do both processes lead to the same final state. If $t'$ 
has the appropriate sign (which is the case for the KLM \cite{Sigrist2}) the resulting
interference is ``constructive'' and the FM state has the lowest energy. 
We can see in this example that the motion of the Kondo singlet can stabilize a magnetic phase. 
  
There are different regimes for which the PAM cannot be reduced to a KLM 
by a perturbative approach. One of these situations corresponds to 
the intermediate  valence region: $\epsilon_a \sim E_F$. In this case 
$n^a_i$ is not longer close to one and the $a$ electrons can move.

In Fig.~1, we illustrate the
(one-dimensional) non-interacting bands for the case of interest:
$\epsilon_a$ close to $E_F$ and above the bottom of the $b$ band. If
$|V| \ll |t_b|$, we can identify two subspaces in each band where the
states have either $b$ ($\psi$ subspace) or $a$ 
($\phi$ subspace) character.  The size of
the crossover region around the points where the original
unhybridized $b$ and $a$ bands crossed is proportional to $|V/t_b|$;
that is, it is very small. The creation operators for the 
Wannier orbitals $\psi^{\;}_{{\bf r}\sigma}$ and $\phi^{\;}_{{\bf
r}\sigma}$ associated with each subspace are:

\begin{eqnarray}
\psi^{\dagger}_{{\bf r}\sigma}&=& \frac{1}{\sqrt N}  \Biggl [
\sum_{{\bf k}\in{\bf K}^>} e^{i{\bf k\cdot r}}
\beta^{\dagger}_{{\bf k}\sigma} +\sum_{{\bf k}\in{\bf K}^<} e^{i{\bf k\cdot r}} \alpha^{\dagger}_{{\bf
k}\sigma} \Biggr ]
\nonumber \\
\phi^{\dagger}_{{\bf r}\sigma}&=& \frac{1}{\sqrt N} \Biggl [
\sum_{{\bf k}\in{\bf K}^>} e^{i{\bf k\cdot r}}
\alpha^{\dagger}_{{\bf k}\sigma} +\sum_{{\bf k}\in{\bf K}^<} e^{i{\bf k\cdot r}} \beta^{\dagger}_{{\bf k}\sigma}
\Biggr]. \label{phipsi}
\end{eqnarray} 
where $N$ is the number of sites. The subsets ${\bf K}^>$
and ${\bf K}^<$ are defined by: ${\bf K}^>=\{{\bf k}: |u_{\bf k}|\ge |v_{\bf k}| \}$ 
and ${\bf K}^<=\{{\bf k}:|v_{\bf k}| >|u_{\bf k}\}$. Since the $\psi$
and the $\phi$ subspaces are generated by eigenstates of $H_{0}$, it is clear 
that both subspaces can only be mixed by the interacting term $H_U$.
Therefore in the new basis we have:
\begin{eqnarray}
H_{0}=H^{\phi}_{0}+H^{\psi}_{0}=\sum_{{\bf r,r'},\sigma}
\tau^{\phi}_{{\bf r-r'}} \phi^{\dagger}_{{\bf r}\sigma}
\phi^{\;}_{{\bf r'}\sigma}+ \sum_{{\bf r,r'},\sigma}
\tau^{\psi}_{{\bf r-r'}} \psi^{\dagger}_{{\bf r}\sigma}
\psi^{\;}_{{\bf r'}\sigma}, \nonumber
%\label{h00}
\end{eqnarray}
with the hoppings $\tau^{\phi}_{{\bf r}}$ and $\tau^{\psi}_{{\bf r}}$ given 
by the following expressions:
\begin{eqnarray}
\tau^{\phi}_{{\bf r}}= \frac{1}{N}
\Biggl [ \sum_{{\bf k}\in{\bf K}^>}
e^{i{\bf k\cdot r}} E_{{\bf k}}^{-}
+\sum_{{\bf k}\in{\bf K}^<}
e^{i{\bf k\cdot r}} E_{{\bf k}}^{+} \Biggl]
\end{eqnarray}
\begin{eqnarray}
\tau^{\psi}_{{\bf r}}= \frac{1}{N}
\Biggl[ \sum_{{\bf k}\in{\bf K}^>}
e^{i{\bf k\cdot r}} E_{{\bf k}}^{+}
+\sum_{{\bf k}\in{\bf K}^<}
e^{i{\bf k\cdot r}} E_{{\bf k}}^{-} \Biggl].
\end{eqnarray}
The segmented structure of the $\psi$ and the $\phi$
bands introduce oscillations in the hoppings $\tau^{\psi}_{{\bf r}}$
and $\tau^{\phi}_{{\bf r}}$ as a function of the distance 
$|{\bf r}|$.

Because the $U$ term in $H$ involves only the $a$ orbitals, the
matrix elements of $H$ connecting the $\phi$ and $\psi$ subspaces are
small compared to the characteristic energy scales of the problem
(the matrix elements of $H$ within the subspaces). To see this we 
express $a_{{\bf r}\sigma}^\dagger$ as a function of
$\phi^{\dagger}_{{\bf r}\sigma}$ and $\psi^{\dagger}_{{\bf r}\sigma}$
by first inverting Eqs. \ref{alphak} and \ref{phipsi} to find:
\begin{equation}
a_{{\bf r}\sigma}^{\dagger} = \sum_{\bf r'} W_{\bf r-r'}
\phi^{\dagger}_{{\bf r'}\sigma} + w_{\bf r-r'}
\psi^{\dagger}_{{\bf r'}\sigma},
\end{equation}
where the weights $W_{{\bf r}}$ and $w_{{\bf r}}$ are defined by:
\begin{eqnarray}
W_{{\bf r}}= \frac{1}{N} \Biggl [ \sum_{{\bf k}\in{\bf K}^>} 
e^{i{\bf k\cdot r}} u_{\bf k} +\sum_{{\bf k}\in{\bf K}^<} e^{i{\bf k\cdot r}} v_{\bf k} \Biggr ]
\nonumber \\
w_{{\bf r}}= \frac{1}{N} \Biggl [ \sum_{{\bf k}\in{\bf K}^>} e^{i{\bf k\cdot r}} v_{\bf k} + 
\sum_{{\bf k}\in{\bf K}^<} e^{i{\bf k\cdot r}} u_{\bf k} \Biggr ].
\end{eqnarray}
The value of these weights as a function of the distance ${\bf r}$
is plotted in Fig.~2 (for $\epsilon_a=-1.5$, $V=0.1t$ and $t_a=0$).
From Fig.~2 we can see that $W_{{\bf 0}}$ is much larger than 
any other weight. This is so because the $\phi$ orbitals have 
predominantly $a$ character, while the $\psi$ orbitals have mostly $b$
character in the considered region of parameters. 
Therefore we can approximate the creation operator  
$a_{{\bf r}\sigma}^{\dagger}$ by:
\begin{equation}
a_{{\bf r}\sigma}^{\dagger} \approx
\sum_{\bf r'} W_{\bf r-r'} \phi^{\dagger}_{{\bf r}\sigma}
\end{equation}
As a consequence of this approximation, the
$a$ subspace becomes invariant under the application of $H$. In
addition, because $|W^{\;}_{\bf 0}| \gg |W^{\;}_{\bf r \neq 0}|$ (see Fig.~2),
we can establish a hierarchy of terms where the lowest order one
corresponds to a simple on-site repulsion:
\begin{equation}
H^{U}_{eff} = {\tilde U} \sum_{\bf r} n^{\phi}_{{\bf r}\uparrow} n^{\phi}_{{\bf r}\downarrow}
\end{equation}
with $\tilde U=U|W_{\bf 0}|^4$ and $n^{\phi}_{{\bf
r}\sigma}=\phi^{\dagger}_{{\bf r},\sigma} \phi^{\;}_{{\bf r},\sigma}$.
The next order terms, containing three and two $W_{\bf 0}$ factors,
are much smaller and are essentially the same as the intersite
interactions which in the past were added to the Hubbard model to
enhance the ferromagnetism \cite{Vollhardt}.  
\begin{figure}[tbp]
\begin{center}
\vspace{-0.0cm} \epsfig{file=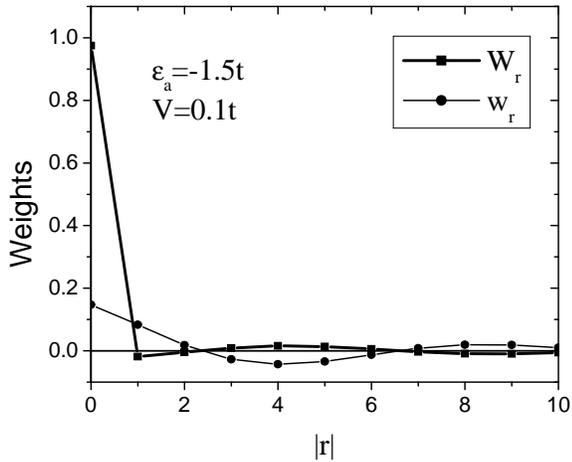,height=70mm,angle=-90}
\end{center}
\vspace{-0.5cm} \caption{Weights $W_{\bf r}$ and $w_{\bf r}$ as 
a function of the distance $|{\bf r}|$ for 
a one dimensional system.} \label{fig2}
\end{figure}

Adding $H^{U}_{eff}$ to
$H_{0}$ we get the effective Hamiltonian:
\begin{eqnarray}
H_{eff} &=& \sum_{{\bf r,r'},\sigma} (\tau^{\psi}_{{\bf r-r'}}
\psi^{\dagger}_{{\bf r}\sigma} \psi^{\;}_{{\bf r'}\sigma}+
\tau^{\phi}_{{\bf r-r'}}
\phi^{\dagger}_{{\bf r}\sigma} \phi^{\;}_{{\bf r'}\sigma}) 
\nonumber \\
&+&{\tilde
U} \sum_{\bf r} n^{\phi}_{{\bf r}\uparrow} n^{\phi}_{{\bf
r}\downarrow} 
\label{heff}
\end{eqnarray}
The $\psi$ and $\phi$ orbitals form uncorrelated and correlated
non-hybridized bands: $H_{eff}=H^{\psi}+H^{\phi}$.  For the $\phi$
orbitals we obtain an effective one band Hubbard model with the
peculiar double shell like dispersion relation shown by the thick lines in Fig.~1.

Particularly for $t_{a}=0$, $H^{\phi}$
has a very large density of states in the lower shell of
the $\phi$ band \cite{Vollhardt} which is located near
$\epsilon_a$ . From Fig.~1 it is also clear that
the electrons first doubly occupy the uncorrelated $\psi$ band
states which are below $\epsilon_a$. However, when $E_F$ gets close to $\epsilon_a$,
i.e., the system is in the mixed valence regime, 
the electrons close to the Fermi level go into some of the
correlated $\phi$ states. Then, the interaction term
$H^{U}_{eff}$, combined with the double shell band structure of
$H^{\phi}_{0}$, gives rise to a FM ground state (GS): The electrons
close to $E_F$ spread to higher unoccupied ${\bf k}$ states and
polarize, which causes the spatial part of their wave function
to become antisymmetric, eliminating double occupancy in real space
and reducing the Coulomb repulsion to zero. The cost of polarizing
is just an increase in the kinetic energy proportional to
$\delta_a \sim \hbar v_F \delta_k$, where $v_F$ is the Fermi
velocity and $\delta_k$ is the interval in ${\bf k}$ space in which
the electrons are polarized.

To determine the stability of this unsaturated FM state, we
compare its energy with that of the PM state. If we
were to build a nonmagnetic state with only the states of the
lower $\phi$ shell, we would find a restricted delocalization for
each electron because of the exclusion of the finite set of band
states (${\bf k}$-states) in the upper shell. To avoid the Coulomb 
repulsion $U$ for double occupying a given site, the electrons need 
to occupy all ${\bf k}$-states. This means 
they have to occupy the $\phi$
states in the upper and lower shells. This restricted delocalization is a
direct consequence of Heisenberg's uncertainty principle, and the
resulting localization length depends on the wave vectors, where
the original $b$ and $a$ bands crossed, that define the size
($\Delta{\bf k}$) of each shell. The energy cost for occupying the
$\phi$ states in the upper shell is proportional to the
hybridization gap $\Delta$. Therefore if $U$ is the dominant energy scale
in the problem and $\Delta\gg\delta_a$, the
FM state lies lower in energy than the nonmagnetic state. Under these conditions, the 
effective FM interaction is proportional to the hybridization gap $\Delta$.

This mechanism for ferromagnetism on a lattice is analogous to the intra-atomic
Hund's mechanism polarizing electrons in atoms. In atoms, we also have different 
degenerate (the equivalent of $\delta_a$ is zero) shells separated by an energy gap. If the
valence shell is open, the electrons polarize to avoid
the short range part of the Coulomb repulsion (again reflecting the
Pauli exclusion principle). The energy of an eventual nonmagnetic
state is proportional either to the magnitude of the Coulomb
repulsion or to the energy gap between different 
shells. The interplay between both energies sets the scale of Hund's 
intra-atomic exchange coupling.

\section{Numerical Method} 
 
Our numerical method, the constrained-path Monte Carlo (CPMC) method,
is extensively described and benchmarked elsewhere.\cite{zhang,zhang1} Here
we only discuss its basic features, assumptions and special details
about our use of it.  

In the CPMC method, the ground-state wave function $|\Psi_0\rangle$ is
projected from a known initial wave function $|\Psi_T\rangle$ by a
branching random walk in an over-complete space of Slater determinants
$|\phi\rangle$.  In such a space, we can write $|\Psi_0\rangle =
\sum_\phi c_{\phi} |\phi\rangle$, where $c_\phi>0$.  The random walk
produces an ensemble of $|\phi\rangle$, called random walkers, which
represent $|\Psi_0\rangle$ in the sense that their distribution is a
Monte Carlo sampling of $c_{\phi}/\sum_\phi c_\phi$, that is, a
sampling of the ground-state wave function.

To completely specify $|\Psi_0\rangle$, only determinants satisfying
$\langle\Psi_0|\phi\rangle>0$ are needed because $|\Psi_0\rangle$
resides in either of two degenerate halves of the Slater determinant
space, separated by a nodal plane.  In the CPMC method the fermion
sign problem occurs because walkers can cross this plane as their
orbitals evolve continuously in the random walk. Without {\it a
priori\/} knowledge of this plane, we use a trial wave function
$|\Psi_T\rangle$ and require $\langle\Psi_T|\phi\rangle>0$. The random
walk solves Schr\"odinger's equation in determinant space, but under
an approximate boundary-condition.  This is what is called the
constrained-path approximation.

The quality of the calculation depends on the quality of the trial
wave function $|\Psi_T\rangle$. Fortunately, extensive testing has
demonstrated a significant insensitivity of the results to reasonable
choices: Since the constraint only involves the overall sign of its
overlap with any determinant $|\phi\rangle$, some insensitivity of the
results to $|\Psi_T\rangle$ is
expected \cite{Bonca,zhang,zhang1,mariana1,mariana2,mariana3}.

Besides as starting point and as a condition constraining a random
walker, we also use $|\Psi_T\rangle$ as an importance
function. Specifically, we use $\langle\Psi_T|\phi\rangle$ to bias the
random walk into those parts of Slater determinant space that have a
large overlap with the trial state. For all three uses of
$|\Psi_T\rangle$, it clearly is advantageous to have $|\Psi_T\rangle$
approximate $|\Psi_0\rangle$ as closely as possible. Only in the
constraining of the path does $|\Psi_T\rangle \not= |\Psi_0\rangle$
in general generate an approximation.

We constructed $ |\psi_T\rangle = \prod_\sigma |\phi_T^\sigma\rangle$
from the eigenstates of the non-interacting problem. Because the total
and z-component spin angular momentum, $S$ and $S_z$, are good quantum
numbers, we could choose unequal numbers of up and down electrons to
produce trial states and hence ground states with
$S=S_z=\frac{1}{2}(N_\uparrow-N_\downarrow)$. Whenever possible, we
would simulate closed shells of up and down electrons, as such cases
usually provided energy estimates with the least statistical error,
but because we wanted to study the ground state energy as a function
of $S$, we frequently had to settle for just the up or down shell
being closed. In some cases, the desired value of $S$ could not be
generated from either shell being closed. Also we would select the
non-interacting states so $|\psi_T\rangle$ would be translationally
invariant, even if these states used did not all come from the Fermi
sea.  The use of unrestricted Hartree-Fock eigenstates to
generate$|\phi_T^\sigma \rangle$ instead of the non-interacting eigenstates
generally produced no significant improvement in the results.

In particular, we represented the trial wavefunction as a single
Slater determinant whose columns are the $N_\sigma$ single-particle
orbitals obtained from the exact solution of $H_0$.  We chose the
orbitals with lowest energies given by
$E^{-}_{\bf k}$ and filled them up to a desired number of
electrons $N_e$.
\begin{equation}
 \vert \psi_T \rangle = \prod_{\bf k,\sigma} \alpha_{{\bf
 k},\sigma}^\dagger \vert 0 \rangle,
\end{equation}
where $\vert 0 \rangle$ represents a vacuum for electrons.  Since our
calculations were performed for a less than full lower band, only
states from the lower band were used to construct the trial
wavefunction.

In a typical run we set the average number of random walkers to 400.
We performed 2000 Monte Carlo sweeps before we taking measurements,
and we made the measurements in 40 blocks of 400 steps. By choosing
$\Delta \tau = 0.05$, we reduced the systematic error associated with
the Trotter approximation to be smaller than the statistical error. In
measuring correlation functions, we performed between 20 to 40
back-propagation steps.

\section{Quantum Monte Carlo Results} 

In section III, we have described a new mechanism for 
itinerant ferromagnetism which is present in the 
mixed valence regime for $n>~1/4$. In addition, we mentioned that 
the system is also expected to 
be FM when the $f$ magnetic moments are localized 
($|V| \ll|E_F-\epsilon_a|$) because the effective RKKY coupling  
is negative when the Fermi surface is small ($k_F \sim 0$). In this section
we show that the itinerant and the localized ferromagnetic 
states are continuously connected in the phase diagram of the 
PAM. However, the energy scale of the first state is much larger than 
the RKKY interaction which characterizes the second one. The  
existence of a crossover region between both states could
explain the fact that there are some $f$-electron compounds for which it 
is very difficult to determine whether they are itinerant or localized ferromagnets.

In addition, we will see that the QMC results are consistent with 
the simple picture derived from our effective model (Eq.~[\ref{heff}]). 
According to that picture, the ferromagnetic state in the mixed valence regime
should be similar to a partially polarized non-interacting solution where the 
polarized electrons are the ones occupying the $f$-character orbitals. It is only
in the crossover region of size $V^2/t_b$ (see Fig.~\ref{fig1}), where the orbitals have a
mixed character, that the correlations introduce an appreciable effect. This
effect is the well known Kondo-like singlet correlation between the $d$ and the $f$
electrons. However, it is important to remark that these Kondo singlets only 
exist in an energy interval $V^2/W$ (where $W$ is the $d$ bandwidth), and therefore 
the number of Kondo singlets $N_{KS}$ is much smaller than the number of 
magnetic moments $N_{MM}$: $N_{KS}/N_{MM}\sim V^2/W^2$. This is a simple manifestation 
of the ``exhaustion'' phenomenon described by Nozier\'es \cite{Nozieres,Nozieres2}. Since 
most of the $f$ magnetic moments are ferromagnetically polarized, the role of these 
few Kondo singlets is marginal in our FM solution. Therefore, for the
mixed valence regime with $n >~ 1/4$, the $f$ magnetic moments which are not 
screened by $d$ electrons develop an effective magnetic interaction as a consequence 
of the interplay between the local Coulomb interaction and the particular band structure. 
In other words, the ``collective Kondo state'' which was proposed in the past 
\cite{Nozieres,Nozieres2} is replaced by band ferromagnetism \cite{Fazekas}. 
    
In fact, the nature of local moment compensation in the PAM
differs qualitatively from that in the single impurity Anderson model
\cite{Bonca}. In the PAM, if the ground state is a singlet, then
\begin{equation}
\langle S^z_f(j)^2\rangle = -\sum_i \langle S^z_d(i)S^z_f(j)\rangle
                            -\sum_i \langle S^z_f(i)S^z_f(j)\rangle
\end{equation}
In the impurity model, the last term is absent, and the resulting
expression is the analytic statement of the well-known
Clogston-Anderson compensation theorem that express the compensation
of the f-moment by the conduction electrons. In the PAM, on the other
hand, the last term dominates the first so the f-moment is compensated
largely by correlations induced among themselves. 

\begin{figure}[tbp]
\begin{center}
\vspace{-0.3cm}
\epsfig{file=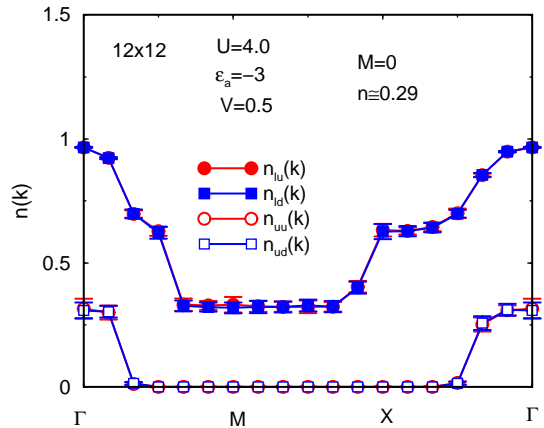,height=70mm,angle=-90}
\vspace{0.0cm}
\end{center}
\caption{Mean value of the occupation numbers of the non-interacting band states
for the paramagnetic $S=0$ solution of the PAM in a two dimensional lattice.  Legends:
$n_{lu}({\bf k})$ represents lower-band occupation number of up-spins,
$n_{ud}({\bf k})$ represents upper-band occupation number of
down-spins, etc.} 
\label{fnk1}
\end{figure}

To understand the nature of the FM solution, we plotted the 
mean occupation number of the quasiparticle operators which 
diagonalize the non-interacting problem (see Eq.~\ref{quasi}). 
This is shown in Fig.~\ref{fnk1} for the lowest energy PM 
($S=0$) state in  a two dimensional cluster of $12 \times 12$  unit cells. 
There we can see that the $d$-character states of
the lower band are close to be doubly occupied. In contrast, the 
$f$-character region has a smaller occupation number ($\sim 0.7$). 
It is remarkable that the populations of the
$f$-character states in the lower and the upper bands are very similar.
This delocalization in the momentum space is a direct consequence of the
tendency to avoid double-occupancy in the real space. This tendency indicates that
the energy increase of the PM state due to the inclusion of $U$
is proportional to the hybridization gap. 

\begin{figure}[tbp]
\begin{center}
\vspace{-0.7cm}
\epsfig{file=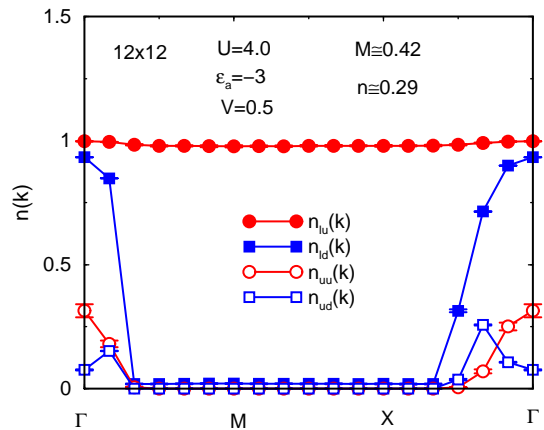,height=70mm,angle=-90}
\vspace{-0.2cm}
\end{center}
\caption{Mean value of the occupation numbers of the non-interacting
band states for the ground state (partially saturated ferromagnet with
$M=S/N=0.43$) of the PAM in a two dimensional lattice.  }
\label{fnk2}
\end{figure}

Fig.~\ref{fnk2} shows the quasiparticle occupation numbers for the 
partially saturated FM ground state. While the $b$-character states of
the lower band are still close to being doubly occupied, the $a$-character
ones are polarized and the occupation number is one for any of them.
In contrast to the PM solution (see Fig.\ref{fnk1}), the occupation number
for the $f$-character states in the upper band is much smaller. This 
difference can be understood in the following way: the polarized $f$-electrons 
can localize in momentum space because the Pauli exclusion principle 
prevents double occupancy in real space (the spatial wave function is 
completely antisymmetric). Therefore the $a$-electrons do not need to occupy
the upper band states and the energy increase due to the repulsive $U$ term is 
proportional to $\hbar v_F \delta_k $. The non-zero amount of electrons 
occupying the center of the upper band (see Fig.~\ref{fnk2}) is related to 
the crossover region for which the $a$ and the $b$ character of the states
are comparable. Since the electrons occupying these states are not polarized,
the effect of the Coulomb repulsion $U$ is the transfer of spectral weight from 
the lower to the upper band to avoid double occupancy in real space. 

\begin{figure}[tbp]
\begin{center}
\vspace{-0.7cm}
\epsfig{file=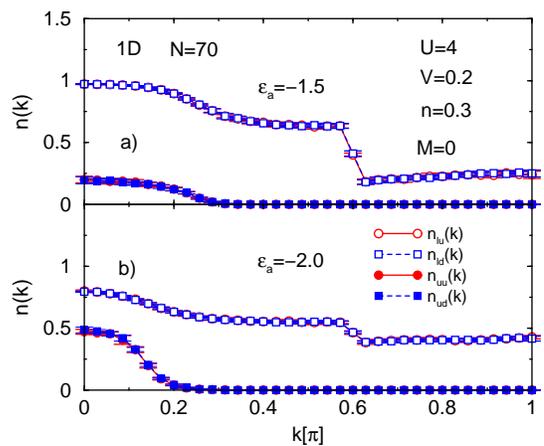,height=70mm,angle=-90}
\vspace{-0.2cm}
\end{center}
\caption{Mean value of the occupation numbers of the non-interacting band states
for the paramagnetic $S=0$ solution of the PAM in a chain.} 
\label{fn1d}
\end{figure}

\begin{figure}[tbp]
\begin{center}
\vspace{-0.7cm}
\epsfig{file=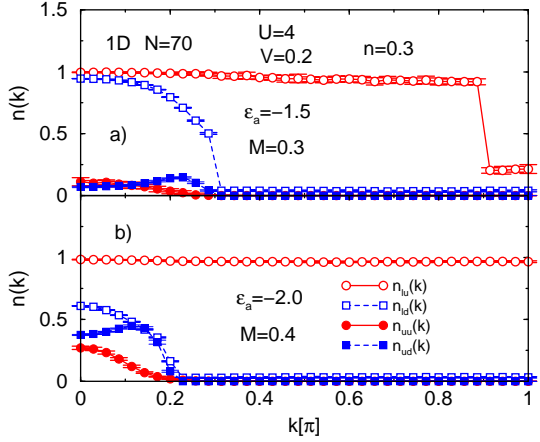,height=70mm,angle=-90}
\vspace{-0.3cm}
\end{center}
\caption{Mean value of the occupation numbers of the non-interacting band states
for the ground state (partially saturated ferromagnet with $M=S/N=0.3$) of the PAM
in a chain.} 
\label{fn1d2}
\end{figure}

We can see from Figs.~\ref{fn1d} and \ref{fn1d2} that a similar behavior 
is obtained for one dimensional systems. As we explained in section III,
the mechanism for this ferromagnetism works in any finite dimension.  
Notice in Fig.~\ref{fn1d} that there is a jump in the occupation number 
of the lower band  as a function of ${\bf k}$. The inverse of this 
jump is proportional to effective mass of the quasiparticles 
of the pramagnetic solution. When $\epsilon_a$ decreases, the 
system evolves into a state where the $a$ electrons are localized. 
This evolution is reflected in the decrease of the jump and the 
consequent increase of the effective mass.

\begin{figure}[tbp]
\begin{center}
\vspace{-0.7cm}
\epsfig{file=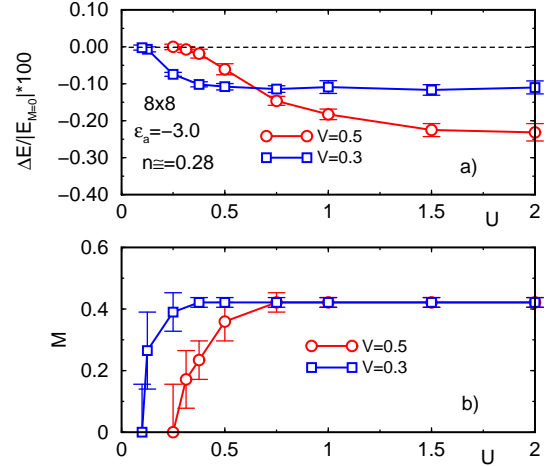,height=70mm,angle=-90}
\vspace{-0.3cm}
\end{center}
\caption{a) Energy difference between partially polarized FM ground state
and the lowest energy paramagnetic state as a function of $U$. b) Magnetization
as a function of $U$.} 
\label{evsU}
\end{figure}

For $U=0$ the system is a PM metal. Therefore there must be
a critical value of the on site repulsion $U_c$ which separates the
PM from the FM region.  The value obtained for $U_c$ is:
$U_c\sim 0.25 \sim \Delta=0.242$ for $V=0.5$ and $U_c\sim 0.11 \sim \Delta=0.0981$ for
$V=0.31$ (see Fig.~\ref{evsU}).  This is also in agreement with the
mechanism described in section III. If $\Delta \gg \delta$ and $U$
becomes larger than the hybridization gap $\Delta$, the system evolves
into a FM state to avoid the double occupancy without an increase in
the kinetic energy proportional to the hybridization gap.

\begin{figure}[tbp]
\begin{center}
\vspace{-0.7cm}
\epsfig{file=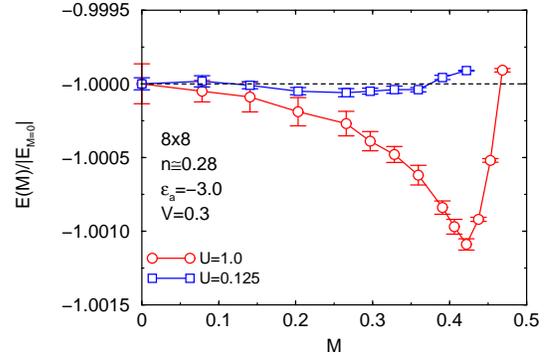,height=70mm,angle=-90}
\vspace{-0.3cm}
\end{center}
\caption{Minimum energy as a function of the magnetization.} 
\label{evss}
\end{figure}

According to Figs.~\ref{evsU}b and \ref{evss} the magnetization seems to increase 
gradually when $U$ is increased beyond its critical value. This behavior 
suggests that the FM transition as function of $U$ is of second order. 
If this is so, the properties of the PM Fermi liquid which is obtained
for $U<U_c$ should be strongly affected by the FM fluctuations
when $U$ approaches $U_c$. It is known that the effective mass of the 
quasi-particles diverges in the approach to a zero temperature ferromagnetic 
instability \cite{Brinkman,Vollhardt2,Beal,Moriya}. In other words, the 
behavior of the PM Fermi liquid cannot be understood by analogy 
with the one impurity problem. The Kondo temperature, which is the characteristic 
energy scale of the one impurity problem, is replaced by a new Fermi temperature 
which is dominated by the ferromagnetic fluctuations and goes to zero when 
$U$ approaches to $U_c$ from below. 

\begin{figure}[tbp]
\begin{center}
\vspace{-0.7cm}
\epsfig{file=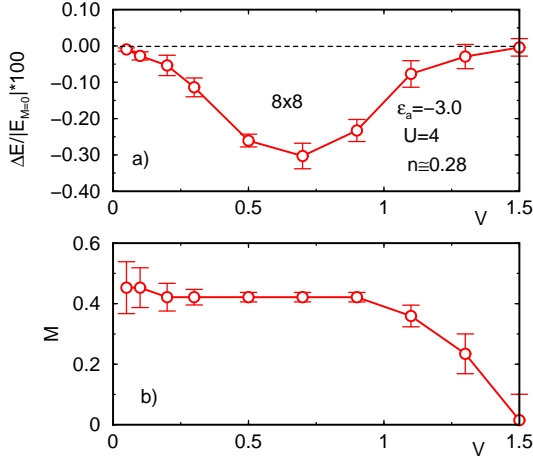,height=70mm,angle=-90}
\vspace{-0.3cm}
\end{center}
\caption{a) Energy difference between the FM ground state and the lowest energy 
paramagnetic state as a function of $V$. b) Magnetization
as a function of $V$.} 
\label{evsv}
\end{figure}

Another relevant parameter for the FM solution is the hybridization 
$V$. For $V=0$ and $\epsilon_a=E_F$ there is a complete spin degeneracy for
the electrons occupying the localized $a$ orbitals. By increasing $V$, we are 
simultaneously changing the Fermi velocity $v_F$ and the hybridization gap $\Delta$. 
For small values of $V$, $\Delta$ is much larger than $\delta=\hbar v_F \delta_k$ 
in the region under consideration. For this reason, a non-zero value of $V$ 
removes the original spin degeneracy stabilizing the partially polarized FM solution
(see Fig.~\ref{evsv}). When $V$ is larger than $t$, the two relevant energy scales, 
$\Delta$ and $\delta$, become of the same order and the partially polarized FM 
is replaced by a PM phase. In the unrealistic large $V$ limit 
($|V| \gg |t_b|,U,|\epsilon_a|$), the ground state consists of local Kondo 
singlets moving in a background of localized spins \cite{nos1}.

% Ferromagnetism in the 1D case? 

\begin{figure}[tbp]
\begin{center}
\vspace{-0.7cm}
\epsfig{file=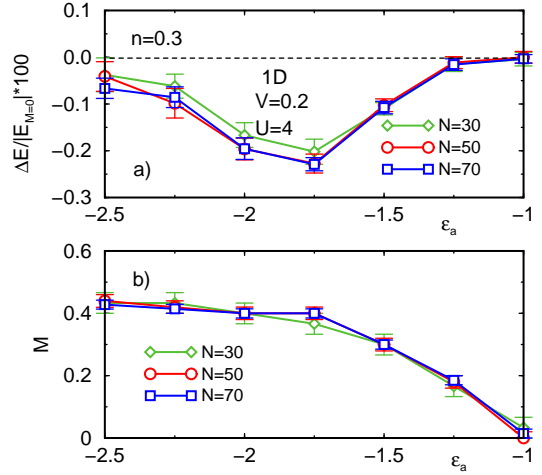,height=70mm,angle=-90}
\vspace{-0.1cm}
\end{center}
\caption{a) Energy difference between FM ground state and  the lowest energy paramagnetic state
as a function of $\epsilon_a$ for a one dimensional system. b) Magnetization as a function of $\epsilon_a$.} 
\label{evsef}
\end{figure}

Finally, the most sensitive parameter for the stabilization of the 
FM state is the difference $|\epsilon_a-E_F|$. In Fig.~\ref{evsef},
we show the energy difference $|\Delta E|$ between the FM ground state and the 
lowest energy PM ($S=0$) state as a function of $\epsilon_a$.
When $\epsilon_a$ is considerably smaller than $E_F$, the $a$ electrons
are localized and the magnetism is dominated by the RKKY interaction.
The order of this interaction is $V^4$. This gives the small value
of $|\Delta E|$ when the $a$ levels are below the bottom of the 
conduction band $\epsilon_a < -2D|t_b|$. The most stable region for the 
FM state (maximum value of $|\Delta E|$) starts when $\epsilon_a$ reaches the
Fermi level. For the case of Fig.~\ref{evsef}, this occurs at $\epsilon_a \sim -1.9t$.
Again this result is in agreement with the mechanism described in section III. 
If we continue increasing the value of $\epsilon_a$, the number of $a$ electrons 
decreases and the magnetization is consequently reduced (see Fig.~\ref{evsef}b). 
Finally, when $\epsilon_a$ is no longer close to the bottom of the conduction band,
$\delta_a$ becomes comparable to $\Delta$ and the ferromagnetism disappears. 

\begin{figure}[tbp]
\begin{center}
\vspace{-0.7cm}
\epsfig{file=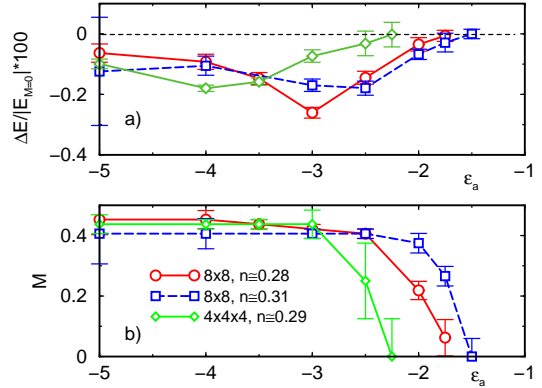,height=70mm,angle=-90}
\vspace{-0.1cm}
\end{center}
\caption{a) Energy difference between FM ground state and  the lowest energy paramagnetic state
as a function of $\epsilon_a$ for two and three dimensional systems. b) Magnetization as a function of $\epsilon_a$.} 
\label{evsef2}
\end{figure}

Fig.~\ref{evsef2} shows the $\epsilon_a$ dependence of $|\Delta E|$ for 
two and three dimensional clusters. As in the one dimensional case, 
the stablity of the FM phase increases when the system appraches the 
mixed valence regime. If $\epsilon_a$ is further increased, the 
magnetization goes to zero in a smooth way indicating that the 
associated quantum phase transition is of second order.

\begin{figure}[tbp]
\begin{center}
\vspace{-0.7cm}
\epsfig{file=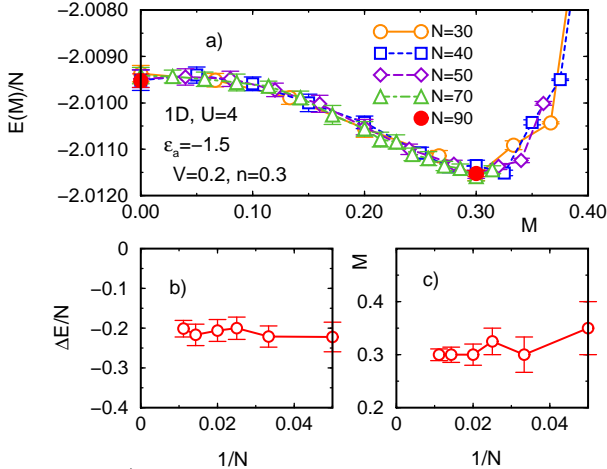,height=80mm,angle=-90}
\vspace{-0.3cm}
\end{center}
\caption{a) Scaling of the energy difference between the FM ground state  
and the lowest energy paramagnetic state for different chain lengths. b)
Scaling of the magnetization per site.} 
\label{scaled1}
\end{figure}

\begin{figure}[tbp]
\begin{center}
\vspace{-0.7cm}
\epsfig{file=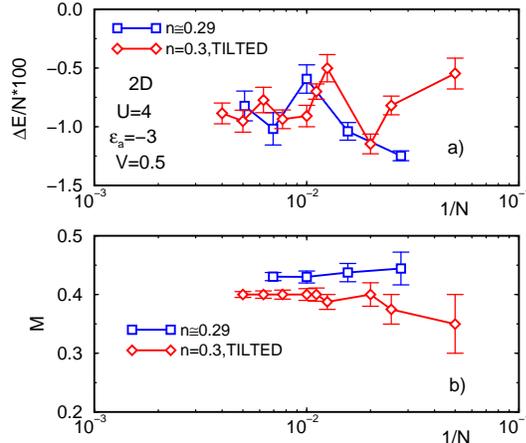,height=70mm,angle=-90}
\vspace{-0.3cm}
\end{center}
\caption{a) Scaling of the energy difference between the FM ground state  
and the lowest energy paramagnetic state for two dimensional systems. b)
Scaling of the magnetization per site.} 
\label{scaled2}
\end{figure}

In Figs.~\ref{scaled1} and \ref{scaled2} we show the scaling of $\Delta E$ and the 
magnetization per site in one and two dimensional systems. The $1/N$ extrapolation
for the one dimensional systems indicates that the FM ground state is stable in the 
thermodynamic limit. The extrapolated value for the magnetization is $\sim 0.3$, i.e.,
only a fraction of the $f$ electrons is polarized. The size effects are stronger
in two dimensional systems. In Fig.~\ref{scaled2}, we show two different cases: the 
circles correspond to a sequence of tilted suqare clusters which allows us to fix the 
concentration in $n=0.3$; the squares correspond to a sequence of untilted square clusters 
for which the value of the concentration is the closest to $n=0.29$. Despite the 
considerable size effects, these results indicate that the FM state is stable in
the thermodynamic limit. In this case, the extrapolated magnetization is close to 0.4. 

%-Comparison between the Magnetization 
%as a function of doping calculated with PAM 
%and with the effective model

%-Scaling of Energy and Magnetization

% Finite size results in 3D?

Meyer and Nolting \cite{meyer3} have also found a FM solution for the PAM
in a similar region of parameters using DMFT. However it is important to remark that
the mechanism for ferromagnetism described in section III only works in finite
dimension because the volume of the upper shell relative to the volume of the 
lower one goes to zero when the dimension goes to infinity. Therefore, the reason
why a FM state is obtained with DMFT should be different. This is also reflected 
by the fact that the energy scale ($T_c$) of the FM solution found with DMFT for the localized 
regime is larger than the one for the mixed valence regime (region IV of Ref.~\cite{meyer1}).
This behavior is opposite to our result (see Fig.~\ref{evsef}). 

\section{Experimental Consequences}

\subsection{Cerium Compounds}

During the last few years new experimental results have confirmed 
that there are Ce based compounds which cannot be treated as 
typical Kondo systems. For instance, CeRh$_3$B$_2$ has a
very high FM ordering temperature ($T_c = 115^{\circ}K$) which sicks out
from a localized $4f$-electron description \cite{Berger}. In addition 
the alloy series Ce(Rh$_{1-x}$Ru$_x$)$_3$B$_2$ \cite{Berger} and  
La$_x$Ce$_{1-x}$Rh$_3$B$_2$ \cite{shaheen} exhibit many unusual 
characteristics which require a new macroscopic description 
with respect to the competition among classical Kondo versus RKKY 
interactions \cite{Doniach0}.

Absorption edge spectroscopy measurements  of Ce(Rh$_{1-x}$Ru$_x$)$_3$B$_2$
for different values of $x$ indicate that the stoichiometric compound 
CeRh$_3$B$_2$ is in the mixed valence regime (fluctuating between 
the $4f^1$ and the $4f^0$ configurations) \cite{Berger}. After doping with 
Ru, there is strong transfer of weight from the $4f^1$ line to the  $4f^0$
structure. This change can be understood in the context of the PAM 
if we take into account that 
the volume of the system decreases when Ce is replaced by Rh \cite{shaheen}.
In this situation the width of the conduction band increases and some $f$ 
electrons are transferred to $d$ character orbitals (see Fig.\ref{fig1}).
According to our results this change must decrease the value of the zero 
temperature magnetization and the Curie temperature $T_c$.  By increasing the 
doping level we can reach a situation where most of the $f$ electrons that 
were polarized in the stoichiometric compound are transferred to the 
$d$ character orbitals and the zero temperature  magnetization is very small.
In this limit the system should have some PM states very close
to the Fermi level (see Fig.~\ref{fig1}). When the energy difference 
between the Fermi level and these PM states becomes smaller than
$T_c$, the magnetization can increase with temperature because 
the electrons which are occupying the PM states near the Fermi level
are thermally promoted to the $f$ character states which are above the Fermi level. 
In this process, the electrons are polarized because of the mechanism discussed above. 
This explains the finite temperature peak in the magnetization of
Ce(Rh$_{1-x}$Ru$_{x}$)$_3$B$_2$ (for $x$ between 0.06 and 0.125) \cite{malik,Berger}
that suggests an ordered state with high entropy. The source of the large entropy is thus
associated with charge and not with spin degrees of freedom which
is why a state with larger $M$ has a higher
entropy. From this analysis we predict that the integral of the entropy
below $T_c$, which can be extracted from the specific heat measurements,
contains a considerable contribution from the {\it the charge} degrees of freedom.

When Ce is replaced by La, the volume of the system increases
\cite{shaheen} and the magnetic moments become more localized. In this case,
the weight in the absorption edge spectroscopy is transferred from the 
$4f^{0}$ structure to the $4f^{1}$ line. Again this can be understood 
if we take into account that the width of the conduction band decreases 
in this case and the electrons are transferred from the $d$ to the 
$f$ character orbitals (see \ref{fig1}). In this way the system evolves 
from the itinerant to the localized situation ($\epsilon_f=\epsilon_a < E_F$).
According to our results (see Fig.\ref{evsef}), this change should increase
the value of the zero temperature magnetization and simultaneously decrease the 
Curie temperature $T_c$ ($|\Delta E|$ is strongly reduced because 
the effective magnetic interaction in the localized limit, $J_{RKKY}$,
is order $V^{4}$ \cite{ours1}). This anomalous behavior has been
experimentally observed by Shaheen et al \cite{shaheen} in La$_x$Ce$_{1-x}$Rh$_3$B$_2$.

\begin{figure}[tbp]
\begin{center}
\vspace{-0.7cm}
\epsfig{file=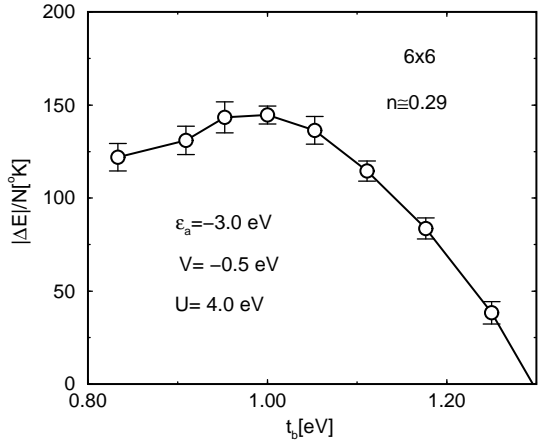,height=70mm,angle=-90}
\vspace{-0.3cm}
\end{center}
\caption{a) Energy difference per site between the FM and PM
states as a function of $t_b$.} 
\label{press}
\end{figure}

We can also connect our mechanism with the hydrostatic pressure
dependence of $T_c$. To do this we calculated $|\Delta E|/N$ by
the QMC method as function of increasing $t_b$ (Fig.~\ref{press}a). Here we
are assuming that the main effect of the hydrostatic pressure is
to increase $t_b$ and to leave the other parameters unchanged. The
order of magnitude of $|\Delta E|/N$, which should be proportional
to $T_c$, and its qualitative behavior in Fig.~3a are in good
agreement with the experimental results for CeRh$_3$B$_2$
\cite{Cornelius}. We see from Fig.~3a that for the
itinerant FM case, $|\Delta E|$/N 
is of the order of 100$^{\circ}$K. This scale is much larger than the
magnitude of the RKKY interaction \cite{degennes} ($\sim
1^\circ$K) which is commonly used to explain the origin of the
magnetic phase when the $a$ electrons are localized.  We also find
that the FM state appears close to quarter filling and disappears
for $n$ close to $3/8$.

\subsection{Uranium Compounds}

The FM uranium monochalcogenides US, USe and UTe are semi-metals 
with large Curie temperatures of $T_c=180K$, $160K$ and $108K$ and 
ordered moments of 1.5, 2.0 and 2.2$\mu_B$ respectively \cite{santini}.
Most of the magnetic properties of these systems are still unexplained. 
The purpose of this subsection is to argue that the new mechanism for 
ferromagnetism introduced above is a good candidate to explain some of the 
mysteries related to these compounds. 

Erd\"os and Robinson \cite{erdos} suggested that the uranium monochalcogenides 
are mixed valence systems. This suggestion was reinforced by measuring the 
Poisson ratio as a function of the chalcogen mass with low
temperature ultrasonic studies on USe and UTe \cite{neuenschwander}. 
The coexistence of an intermediate valence regime and ferromagnetism is one
of the unexplained properties of these compounds as it 
is recognized in Ref.~\cite{santini}. According to the traditional picture
\cite{Doniach0}, the system should behave as a nonmagnetic collective Kondo state
in the intermediate valence regime. In contrast to this picture, our results show that 
a partially saturated ferromagnetic state is stabilized in the mixed valence regime.
This can explain the first striking property of the uranium monochalcogenides.

The other unusual property of these compounds is the shape of the magnetization 
curve versus temperature which has a maximum below $T_c$ \cite{erdos}.
Again this is a property which can be easily explained (see the subsection about
the Ce compounds) within the context of the PAM. In addition, the order of magnitude
of the Curie temperature of these compounds coincides with the energy scale 
obtained from the PAM for the intermediate valence regime. 

The uranium monochalcogenides, like the Ce based compounds above described,
exhibit a non-monotonic behavior for the $T_c$ as a function of pressure \cite{Cornelius}. 
Fig.~\ref{press} shows that this behavior can also be explained with the PAM. 
Notice however that the non-monotonic behavior shown in Fig.~\ref{press} has 
nothing to do with a competition between Kondo and RKKY interactions.

Finally, the spin wave dispersions of these compounds also present some anomalies.
For instance, neutron scattering experiments on a single-domain UTe crystal 
(\cite{lander1,lander2}) show that for wave vectors $q$ perpendicular to the 
ordered moment the excitations become more damped with increasing $q$. In US 
only a broad continuum of magnetic response is observed \cite{buyers}. Damped and 
unpolarized spin waves are observed in USe \cite{duplessis,holden,hughes}. 
These properties indicate that the itinerant character of the $f$ electrons 
is essential to have a good description of the magnetic excitations.

\subsection{Transition Metals}

Even though the  transition metals are the most well studied itinerant ferromagnets,
the ultimate reason for the stabilization of the FM phase is still unknown. Since the 
minimal correlated model (Hubbard Hamiltonian) proposed to describe these systems 
does not seem to have a FM solution, it is reasonable to ask whether an extension of this minimal 
model, including more than one band, is necessary and enough to stabilize the 
FM solution. The correlated $3d$ band of the transition metals is hybridized 
with weakly correlated and dispersive $s$ and $p$ bands. This situation is similar to the case
already described for the $f$ electron compounds. Therefore, it is natural to
ask if there is a connection between the itinerant ferromagnetism of the 
$f$ and the $d$ electron compounds. Notice that the order of magnitude 
of the $T_c$ is the same. Following the same motivation and using DMFT, 
Schwieger and Nolting \cite{Schwieger} concluded that the $d$-band ferromagnetism can 
be stabilized when the hybridization between both bands is small. However, these authors
also find a FM solution for the one band problem.

\begin{figure}[tbp]
\begin{center}
\vspace{-0.7cm}
\epsfig{file=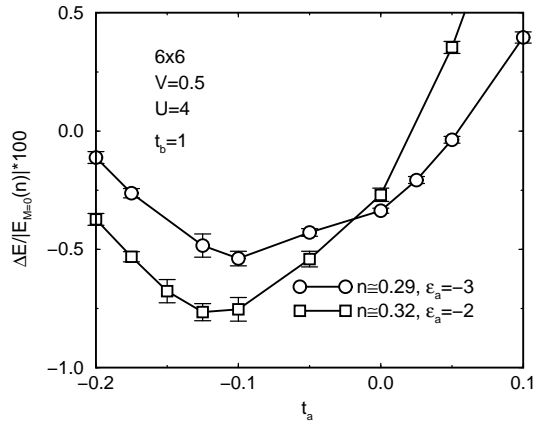,height=70mm,angle=-90}
\vspace{-0.3cm}
\end{center}
\caption{Influence of the hopping $t_a$ on the FM state.} 
\label{press2}
\end{figure}

\begin{figure}[tbp]
\begin{center}
\vspace{-0.7cm}
\epsfig{file=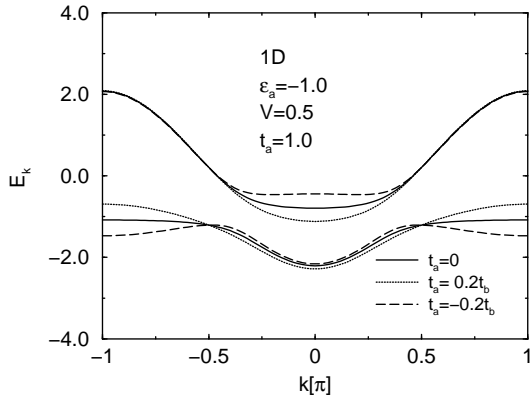,height=70mm,angle=-90}
\vspace{-0.3cm}
\end{center}
\caption{Band structure for different values of $t_a$.} 
\label{ekta}
\end{figure}

In the case of the transition metals, the dispersion of the narrower band ($3d$) cannot
be neglected. For this reason, we studied the stability of the FM solution as 
a function of $t_a$. We can see from Fig.~\ref{press2} that the FM phase is even more
stable for $t_a \sim -0.1$ than for $t_a=0$ and becomes unstable
for $t_a \sim 0.05$. The reason for this asymmetric behavior is easy
to understand in terms of the variation of $\delta_a$: If $t_a$ is
negative, then the effect of $t_a$ on the dispersion of the $\phi$
band is opposite to that of the hybridization $V$ (see Fig.~\ref{ekta}). When $t_a\sim
-0.1$, we get, for the given $\epsilon_a$ and $V$, the minimum value for $\delta_a/\Delta$ 
and therefore the most stable FM case. When we depart from this value of $t_a$,
$\delta_a/\Delta$ increases, $|\Delta E|$ decreases, and the FM state becomes less stable.

This result indicates that the hybridization between bands 
can play a crucial role for the ferromagnetism of the iron group.
In other words, the ferromagnetism of the transition metals 
can originate, at least in part, in the interplay between the correlations 
and the particular band structure and not solely in the intra-atomic 
Hund's exchange \cite{Van Vleck}. 

\section{Conclusions} 
 
We introduced a novel mechanism for itinerant ferromagnetism which 
is present in a simple two band model consisting
of a narrow correlated band hybridized with a dispersive 
and uncorrelated one. The picture just presented, combined with our
previous results \cite{ours1}, allows a reconciliation of the
localized and delocalized ferromagnetism pictures painted by
Heisenberg \cite{Heisenberg} and Bloch \cite{Bloch}.  
The hybridization between bands and the particular band structure play 
a crucial role in this mechanism because they generate a multi-shell
structure for the correlated orbitals. This structure, when combined 
with the a local Coulomb repulsion, favors a ferromagnetic state. The 
mechanism is analogous to the one which generates the atomic Hund interaction.
In this sense, this is a generalization to the solid of the atomic Hund's rule.
The mechanism works in any finite dimension.

The determination of a minimal model to explain 
the metallic ferromagnetism of highly correlated systems has
been the object of intense effort during the last forty years.
The results presented in this paper suggest that the PAM is a  minimal 
Hamiltonian which can explain the itinerant ferromagnetism 
without including any explicit FM interaction. 

Another important aspect of this ferromagnetic solution is its 
mixed valence character. According to the traditional picture \cite{Doniach0},
the mixed valence regime should be a PM Kondo state. 
The appearance of a ferromagnetic instability in this region of 
doping rises some questions about the entire validity of a Kondo-like
description inspired by the one impurity problem. Even the 
PM phase obtained for $U<U_c$ is strongly influenced by 
the proximity to a FM instability \cite{Brinkman,Vollhardt2,Beal,Moriya}.

We have discussed the relevance of these results for some 
$f$-electron compounds which are itinerant ferromagnets with
high Curie temperatures ($\sim 100^{\circ}K$). In particular,
the are several unusual characteristics of the Ce based compounds Ce(Rh$_{1-x}$Ru$_x$)$_3$B$_2$
and La$_x$Ce$_{1-x}$Rh$_3$B$_2$, and the uranium monochalcogenides US, USe and UTe, which 
can be explained, at least at a qualitative level, with the present mechanism.

We have also considered the case relevant for the iron group where 
the dispersion of the lower band is not negligible. 
The fact that the ferromagnetism is even more stable for finite values 
of $t_a$ when the hoppings of
both bands have opposite signs indicates that our mechanism is
relevant to explain the ferromagnetism of the transition metals, like Ni, where
a correlated and narrow $3d$ band is hybridized with the $4s$ band. 
It suggests that the ferromagnetism in the transition metals 
can originate, at least in part, in the interplay between the correlations 
and the particular band structure,
and not solely in the intra-atomic Hund's exchange \cite{Van Vleck}. 

\acknowledgments 
This work was sponsored by the US DOE. We
acknowledge useful discussions with A. J. Arko, B. H.  Brandow,
J. J. Joyce, G. Lander, J. M. Lawrence, S. Trugman, G. Ortiz, and J. L. Smith.  We thank
J. M. Lawrence for pointing out the experimental work on the Ce
compounds. J. B. acknowledges the support of Slovene Ministry of
Education, Science and Sports and FERLIN.

%              BIBLIOGRAPHY 

\end{multicols} 
 
\end{document}